\documentclass[11pt,twoside]{article}
\usepackage{asp2010}

\def \pv    {P{\sc v}}

\newcommand{\Mdot}{\mbox{$\dot{M}$}}

\newcommand{\beq}{\begin{equation}}
\newcommand{\eeq}{\end{equation}}
\newcommand{\beqa}{\begin{eqnarray}}
\newcommand{\eeqa}{\end{eqnarray}}

\newcommand{\Ha} {H$_{\alpha}$}

\newcommand{\fcl}{\mbox{$f_{\rm cl}$}}
\newcommand{\fvel}{\mbox{$f_{\rm vel}$}}

\resetcounters

\begin{document}

\title{The nature and consequences of clumping in hot, massive star winds}
\author{Jon O. Sundqvist$^1$, Stanley P. Owocki$^1$, and Joachim Puls$^2$
\affil{$^1$University of Delaware, Bartol Research Institute, Newark, Delaware 19716, USA}
\affil{$^2$Universit\"atssternwarte M\"unchen, Scheinerstr. 1, 81679 M\"unchen, Germany\\}}

\begin{abstract}
This review describes the evidence for small-scale structure,
`clumping', in the radiation line-driven winds of hot, massive
stars. In particular, we focus on examining to what extent simulations
of the strong instability inherent to line-driving can explain the
multitude of observational evidence for wind clumping, as well as on
how to properly account for extensive structures in density and
velocity when interpreting the various wind diagnostics used to derive
mass-loss rates.
\end{abstract}

\section{Introduction}

Hot, massive stars possess strong stellar winds driven by line
scattering of the star's intense continuum radiation field. The first
quantitative description of such line driving was given in the seminal
paper by \citet[][`CAK']{Castor75}, assuming a stationary,
homogeneous, and spherically symmetric wind.  But despite the
considerable success of subsequent work assuming such a smooth, steady
outflow \citep[e.g.,][]{Vink00}, the theoretical as well as
observational evidence for a time-dependent, inhomogeneous wind is
irrefutable \citep{Puls08, Hamann08}. This review summarizes the
current status concerning small-scale inhomogeneities, `clumping', in
the winds of single hot stars; for an overview of large-scale wind
inhomogeneities, induced by for example rapid rotation or a strong
magnetic field, see \citet{Puls08}. After first outlining the
theoretical basis in terms of a strong instability inherent to line
driving, we next concentrate on confronting predictions of
corresponding instability models with a sub-set of the large body of
direct and indirect observational evidence for wind clumping. In
particular we focus on the consequences of extensive density and
velocity structure when interpreting various mass-loss diagnostics.

\section{Theory: the line driven instability}
\label{LDI}

Linear stability analysis showed already early on that the line-driven
winds of hot stars should be unstable for velocity perturbations on
scales near and below the Sobolev length $L_{\rm Sob} = v_{\rm
  th}/(dv/dr)$ \citep{Macgregor79, Carlberg80, Owocki84}. The
operation of this strong, intrinsic `line-driven instability' (LDI)
has since been confirmed by direct numerical modeling of the
time-dependent wind \citep{Owocki88, Feldmeier95, Dessart05}. Such
simulations typically show that the non-linear growth of the LDI leads
to high-speed rarefactions that steepen into strong reverse shocks,
whereby most of the wind material is compressed into dense and
spatially narrow `clumps' that are separated by large regions of much
lower densities. This characteristic structure is the theoretical
basis for our current understanding and interpretation of \textit{wind
  clumping}.

\begin{figure}[h]
\centering
\begin{minipage}{4.0cm}
\resizebox{\hsize}{!}
{\includegraphics[angle=90]{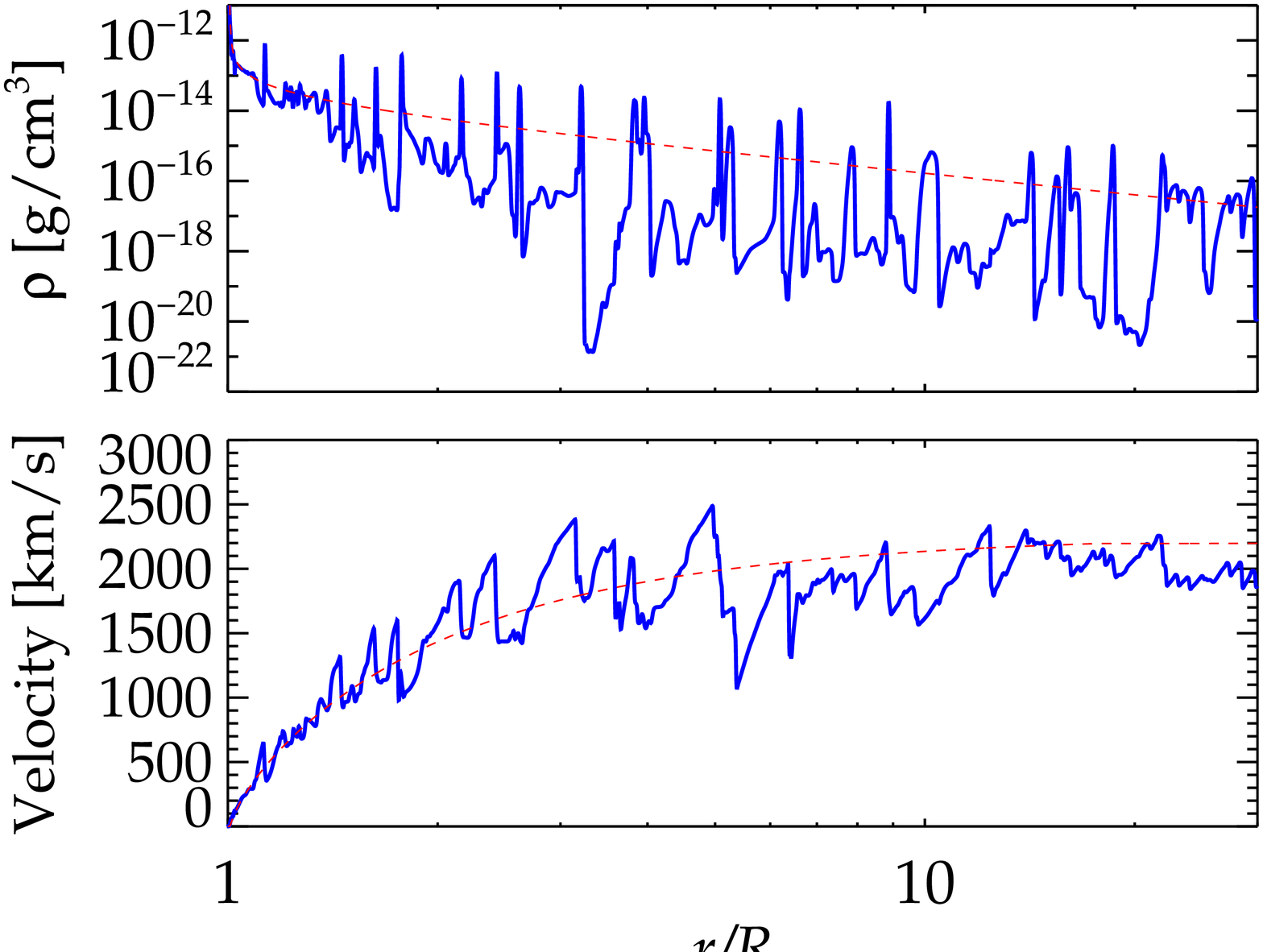}}
\end{minipage}
\begin{minipage}{4.0cm}
\resizebox{\hsize}{!}
{\includegraphics[angle=90]{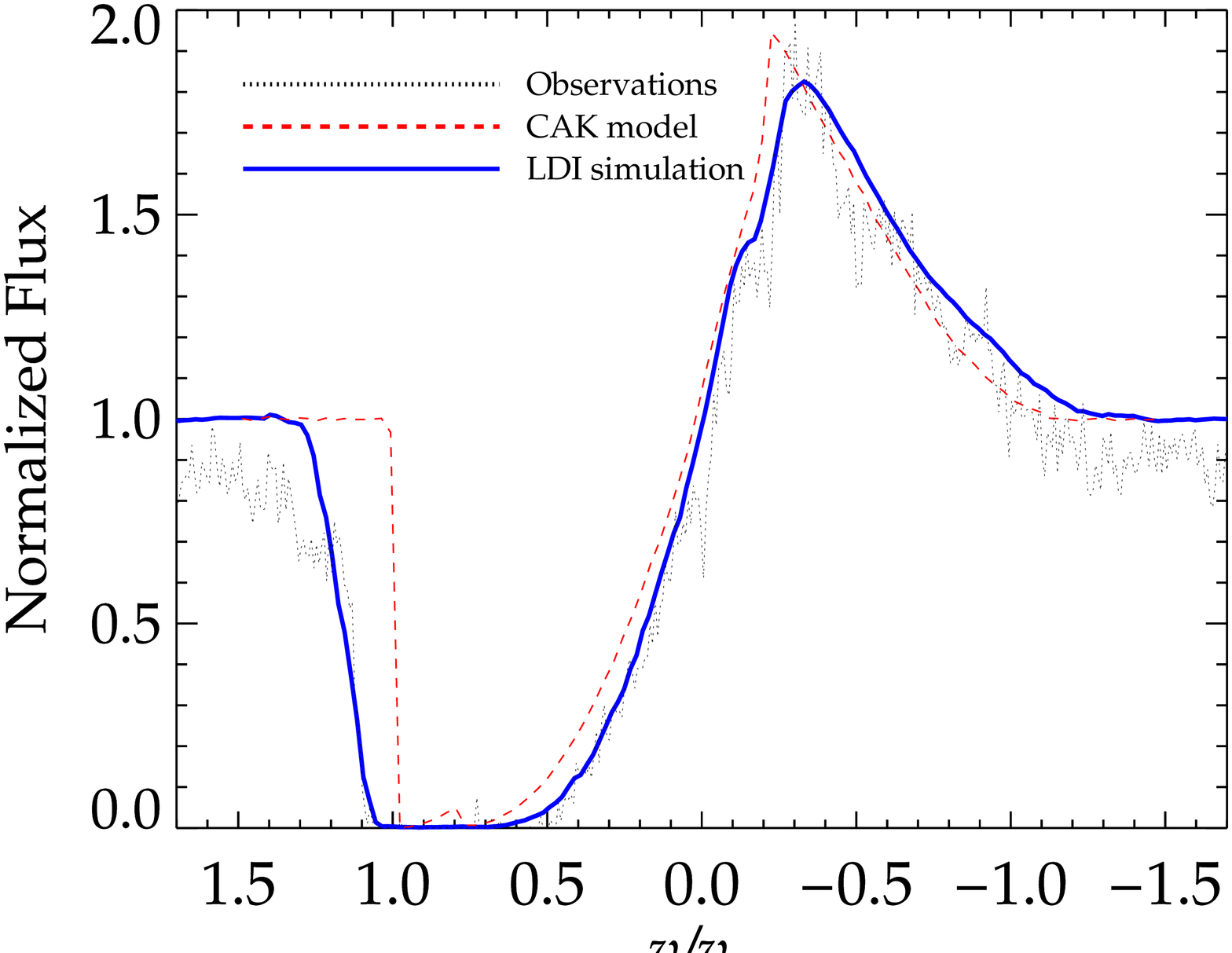}}
\end{minipage}
\begin{minipage}{3.4cm}
\resizebox{\hsize}{!}
{\includegraphics[]{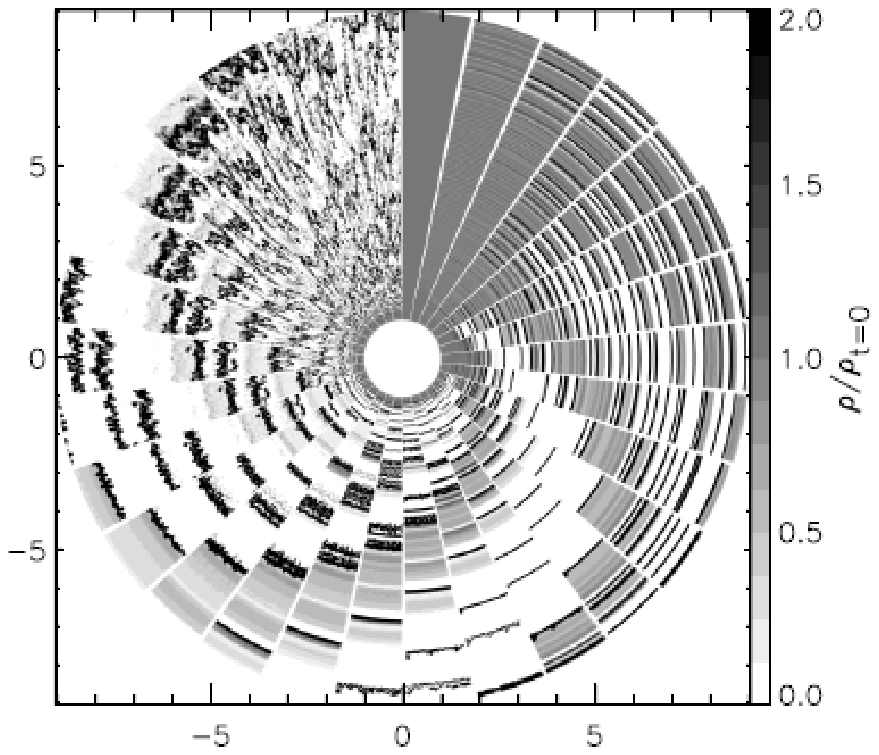}}
\end{minipage}
\caption{\textbf{Left:} Snapshot of density and velocity structure in
  a 1-D LDI simulation (blue), as compared to the smooth CAK start
  model (red, dashed), see text. \textbf{Middle:} Observed IUE spectra
  of the C\,{\sc iv} $\lambda \lambda$1548, 1551 resonance doublet in
  $\zeta$ Pup, as function of line-of-sight velocity of the blue
  component. Overlayed are profiles calculated from CAK and LDI
  models, where the latter synthetic spectrum is computed using an
  extension of the technique developed in \citet{Sundqvist10,
    Sundqvist11}, which takes 1-D snapshots and phase them randomly in
  patches of a parametrized angular size, here of
  1$\deg$. \textbf{Right:} Density contours from a 2-D LDI simulation,
  in which the $\Delta \phi = 12\deg$ wedges represent a clockwise
  time sequence starting from the initial CAK model. Adapted from
  \citet{Dessart03}.}
\label{Fig:LDI}
\end{figure}

The left panel of Fig.~\ref{Fig:LDI} illustrates this typical
structure by plotting density and velocity snapshots of a 1-D LDI
simulation computed from an initial steady CAK model following
\citet{Feldmeier97} \citep[see][for model specific
  details]{Sundqvist11}. The line force is calculated using the
`Smooth Source function' method \citep[SSF,][]{Owocki96}, which allows
one to follow the non-linear evolution of the strong, intrinsic
instability, while also accounting for the stabilizing line drag
effect of the scattered, diffuse radiation field \citep{Lucy84}. The
simulation here further introduces base perturbations from Langevin
turbulence \citep[see][]{Feldmeier97}, which induce structure somewhat
closer to the wind base as compared to simulations with self-excited
structure formation \citep{Runacres02, Runacres05}. But despite these seed
perturbations, the line drag from the SSF force greatly reduces, even
eliminates, the instability near the base, thus allowing the lowermost
wind to remain smooth and steady (Fig.~\ref{Fig:LDI}).
 
The presence of strong, embedded wind shocks in LDI simulations
provides a generally accepted explanation for the X-rays observed from
single hot, massive stars without very strong magnetic fields, as
further discussed by, e.g., \citet{Owocki11} in this volume. In
simulations with self-excited structure though, there is not enough
material going through strong enough shocks to produce the observed
amount of X-rays. But as shown by \citet{Feldmeier97}, introducing
base perturbations such as those considered here enhances the velocity
dispersion of the wind clumps, whereby clump-clump collisions create
regions of relatively dense, hot gas. This then increases the X-ray
emission to levels more comparable to those typically observed.

Because of the computational expense of calculating the line-force at
each time step, LDI simulations have generally been limited to
1-D. More realistically though, thin shell instabilities and
associated effects can be expected to break up the spherical shell
structure into a more complex multi-dimensional structure. First
attempts to construct 2-D LDI simulations have been carried out by
\citet{Dessart03, Dessart05}. Such models typically show that the LDI
first manifests itself as strong density compressions mimicking
corresponding 1-D simulations. But as these initial shell structures
are accelerated outwards, they become disrupted by Rayleigh-Taylor and
thin-shell instabilities, which in the simulations by
\citet{Dessart03} operate all the way down to the azimuthal grid scale
(Fig.~\ref{LDI}, right panel). Moreover, these 2-D simulations have
higher velocity dispersion than corresponding 1-D ones. Similarly as
for the models with base perturbations just discussed then, this 
suggests an increase also in the amount of X-rays produced.

However, it may well be that these initial 2-D models exaggerate
somewhat the level of lateral disruption, as they do not yet properly
treat the lateral component of the diffuse line drag. Presuming this
could damp azimuthal velocity perturbations on scales below the
lateral Sobolev length $L_{\phi} = r v_{\rm th}/v_{\rm r}$
\citep{Rybicki90}, lateral breakup may be prohibited below scales of
order $\Delta \phi \approx L_{\phi}/r \approx v_{\rm th}/v_{\rm r}
\approx 0.5\deg$. Future work should examine this issue by an adequate
incorporation of the lateral line-force into multi-D LDI simulations.

\section{Observations}

\subsection{Line profile variability}

The most direct evidence for clumping in hot star winds comes from
observations of narrow, spectral subpeaks superimposed on broad
optical emission lines in Wolf-Rayet (WR) stars \citep{Moffat88,
  Robert94}. Careful time-monitoring have revealed how these subpeaks,
which trace local wind density enhancements, systematically migrate
from line center toward line edges on time scales that can be
associated with the wind acceleration. Indeed, \citet{Dessart05b}
showed that synthetic spectra calculated directly from LDI models can
match well the observed migration in WR winds; however, the inferred
acceleration scale typically predicts either a more extended
acceleration zone than dynamically expected for line driven winds, or
a much larger core radii than expected for WR stars \citep[see
  also][]{Lepine99}.

While predominately observed in dense WR winds, \citet{Eversberg98}
and \citet{Lepine08} have provided first evidence of emission
substructures also in O-stars, suggesting the phenomenon is universal
in hot star winds. Moreover, \citet{Eversberg98} pointed out that the
density enhancements in $\zeta$ Pup always seemed to appear very near
the stellar surface, thus providing an early indication of clumping
close to the wind base (see Sect.~\ref{stratification}). Further
monitoring of such emission lines formed mainly in the inner wind
would provide crucial information regarding clumping and wind
acceleration in this key region.

\subsection{The absorption troughs of saturated P Cygni lines}

Instability simulations predict extensive structure not only in
density, but also in velocity (Fig.~\ref{Fig:LDI}, left panel). The
most clear-cut evidence for such velocity structure in hot star winds
comes from the extended regions of zero residual flux observed in
saturated UV resonance lines (Fig.~\ref{Fig:LDI}, middle
panel). Already Lucy (1982) suggested that the physical origin of
these `black troughs' is the non-monotonic wind velocity field, which
leads to enhanced back-scattering in multiple resonance zones and
thereby to a systematic reduction of the blue-shifted emission. The
middle panel of Fig.~\ref{Fig:LDI} demonstrates that the velocity
structure predicted by LDI simulations well reproduces the observed
characteristics, whereas the monotonic velocity law of the CAK
comparison model fails to reproduce in particular the blue edge of the
line profile. This clearly indicates the presence of material with
velocities well above the wind terminal speed $v_\infty$ that sets the
maximum blue-shifted absorption in a CAK model.\footnote{The only way
  these features can be modeled within the context of a smooth outflow
  is by introducing a highly supersonic `micro-turbulence'
  \citep{Hamann81}, typically on the order of $\sim$\,5-10\,\% of 
the terminal speed.}

\section{Diagnostics: deriving mass-loss rates from clumped hot star winds}

Due to the rather steady wind base, the average mass-flux in the
simulations described in Sect.~\ref{LDI} is actually not much affected
by the LDI, which may indicate only second order effects on
theoretical mass-loss rates predicted by steady-state models. But let
us now discuss, arguably, \textit{the} key question concerning wind
clumping, namely how the structures in density and velocity affect the
various diagnostics used to infer mass-loss rates from
\textit{observations}.

\subsection{Optically thin clumping}
\label{thin}

Traditionally most diagnostic studies aiming to derive mass-loss rates
have assumed that the wind consists of statistically distributed
\textit{optically thin} clumps embedded in a void background medium,
and neglected any disturbances on the velocity field. The main effect
of such optically thin clumping is that diagnostics having opacities
that scale with the local density squared, such as \Ha~and thermal
IR/radio free-free emission, are stronger than in a smooth wind with
the same mass-loss rate. The scaling invariant is $\sqrt{\fcl}
\dot{M}$, i.e. mass-loss rates derived from \Ha~and smooth wind models
are overestimated with the square-root of the clumping factor $\fcl
\equiv \langle \rho^2 \rangle/ \langle \rho \rangle^2$ in the \Ha~line
forming region. In contrast, diagnostics having opacities that scale
linearly with density, such as UV resonance lines, electron scattering
wings, and bound-free absorption of X-rays, are not directly affected
by optically thin clumping\footnote{They can be indirectly affected
  though, through a modified wind ionization balance due to increased
  recombination rates, e.g.  \citet{Bouret03}.}.

For WR-stars, optically thin clumping has been accounted for since the
pioneering work by \citet{Hillier91}. Simultaneous fitting of electron
scattering wings (scaling linearly with density) and emission peaks
(density-squared) of optical recombination lines indicate $\fcl
\sim$\,4-20, corresponding to mass-loss reductions by factors of
$\sim$\,2-4 \citep[see review by][]{Crowther07}. For O stars, electron
scattering wings are too weak for this technique to be of use. But
combined \Ha/IR/radio studies \citep{Puls06} yield upper limit
mass-loss rates that, similarly, are factors of $\sim$\,2-4 lower than
earlier \Ha~rates based on smooth wind models, and recent X-ray line
attenuation studies confirm such reductions also on an absolute scale
\citep{Cohen10, Cohen11}. However, unsaturated UV resonance lines
often suggest much more drastic reductions \citep{Bouret03, Bouret05,
  Prinja05}. The most prominent example in this respect has been the
analysis of phosphorus {\sc v} (\pv) in 40 Galactic O stars by
\citet{Fullerton06}, who derived values of the mean ionization
fraction times the mass-loss rate, $\langle q_{\rm pv} \rangle
\dot{M}$, that were factors of $10 \dots 100$ lower than corresponding
\Mdot~values derived from smooth models and \Ha/radio emission. Since
state-of-the-art non-LTE model atmospheres that include optically thin
clumping typically predict $\langle q_{\rm pv} \rangle \approx 1$ for
early to mid type O stars, these \pv~results seem to imply extremely
low mass-loss rates, thus challenging the validity of the line-driven
wind theory.
   
\subsection{Optically thick clumping: porosity and vorosity}
\label{thick}

\begin{figure}[h]
\centering
\begin{minipage}{3.5cm}
\resizebox{\hsize}{!}
{\includegraphics[angle=90]{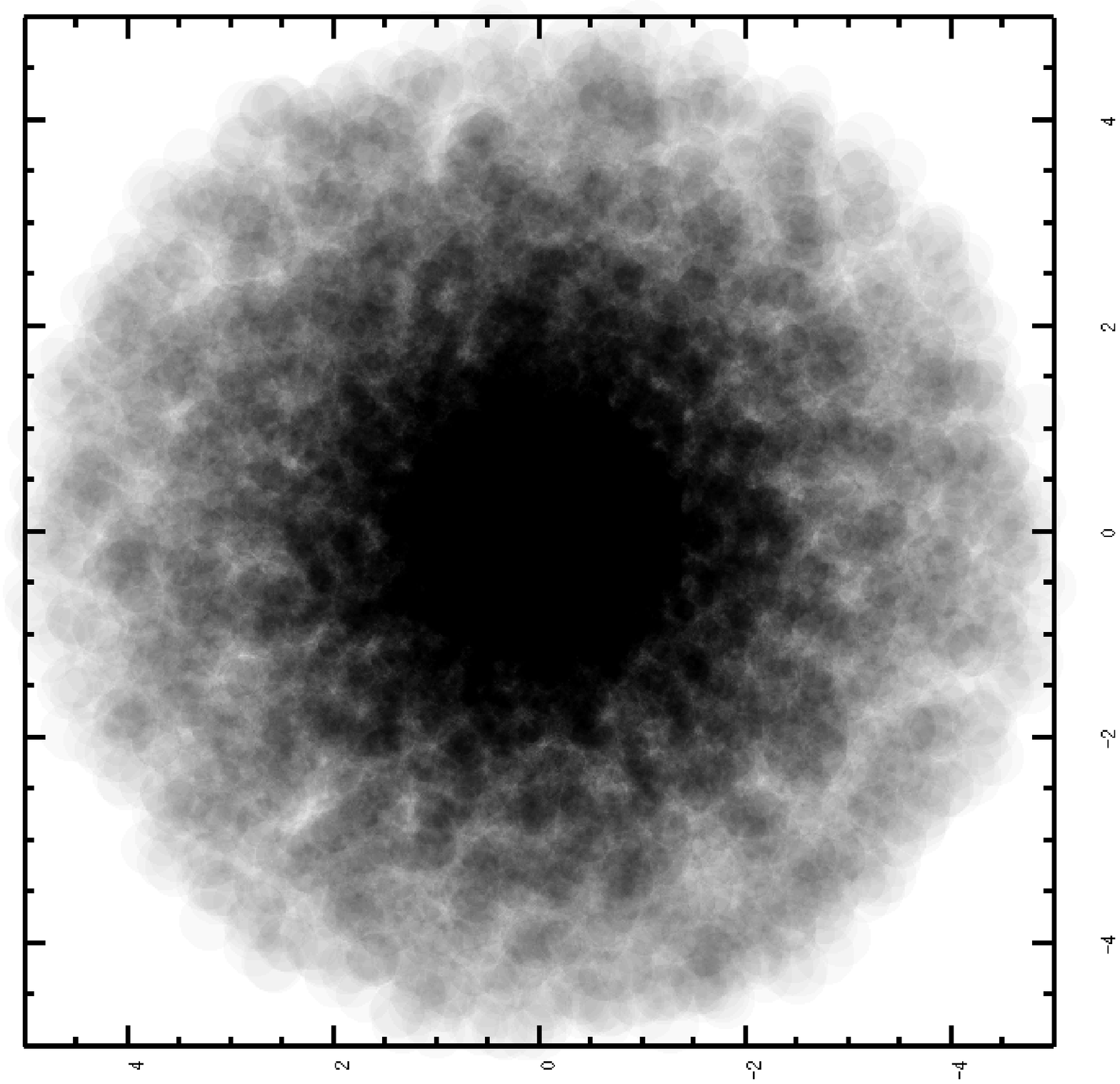}}
\end{minipage}
\begin{minipage}{3.5cm}
\resizebox{\hsize}{!}
{\includegraphics[angle=90]{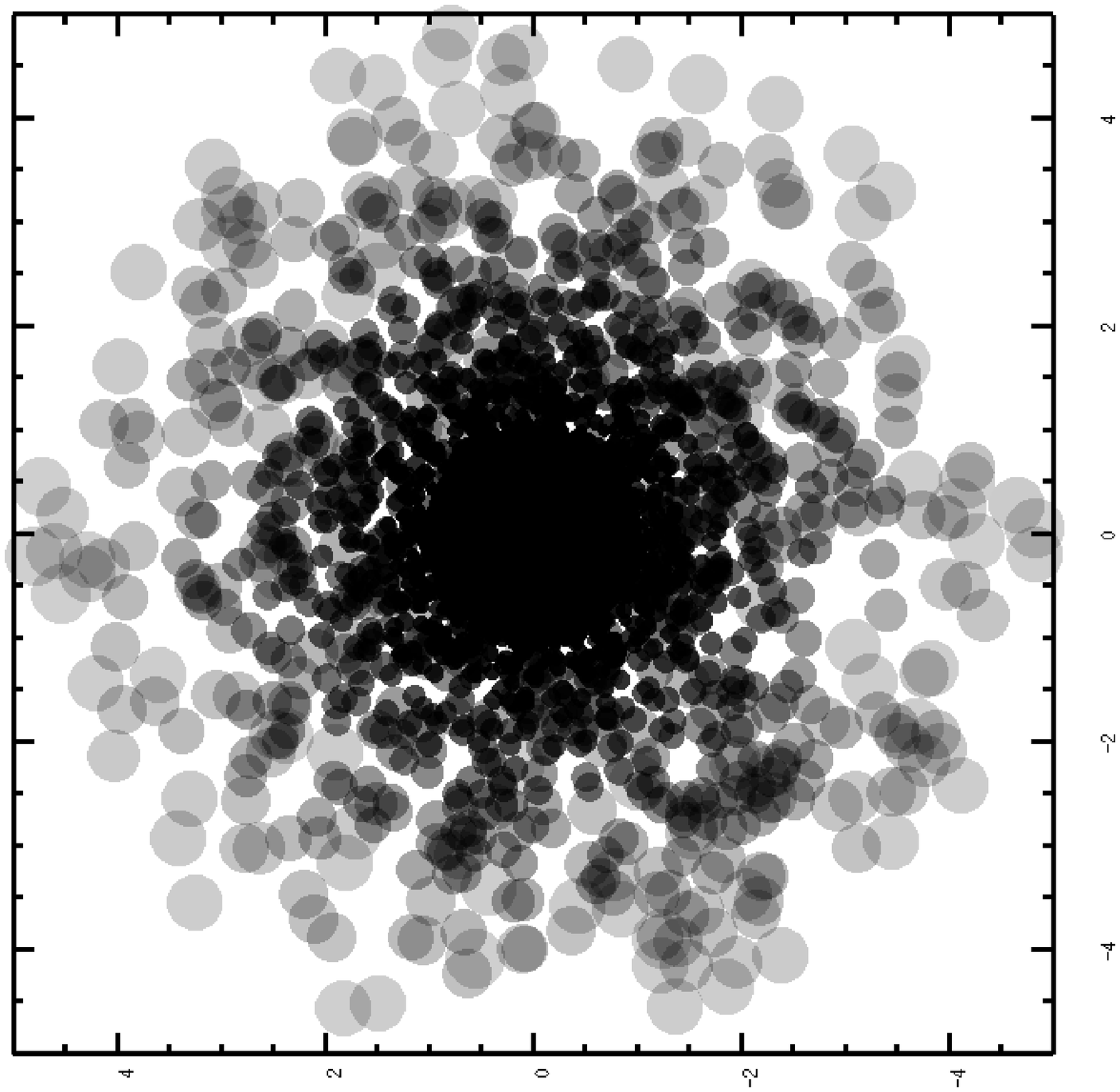}}
\end{minipage}
\begin{minipage}{3.5cm}
\resizebox{\hsize}{!}
{\includegraphics[]{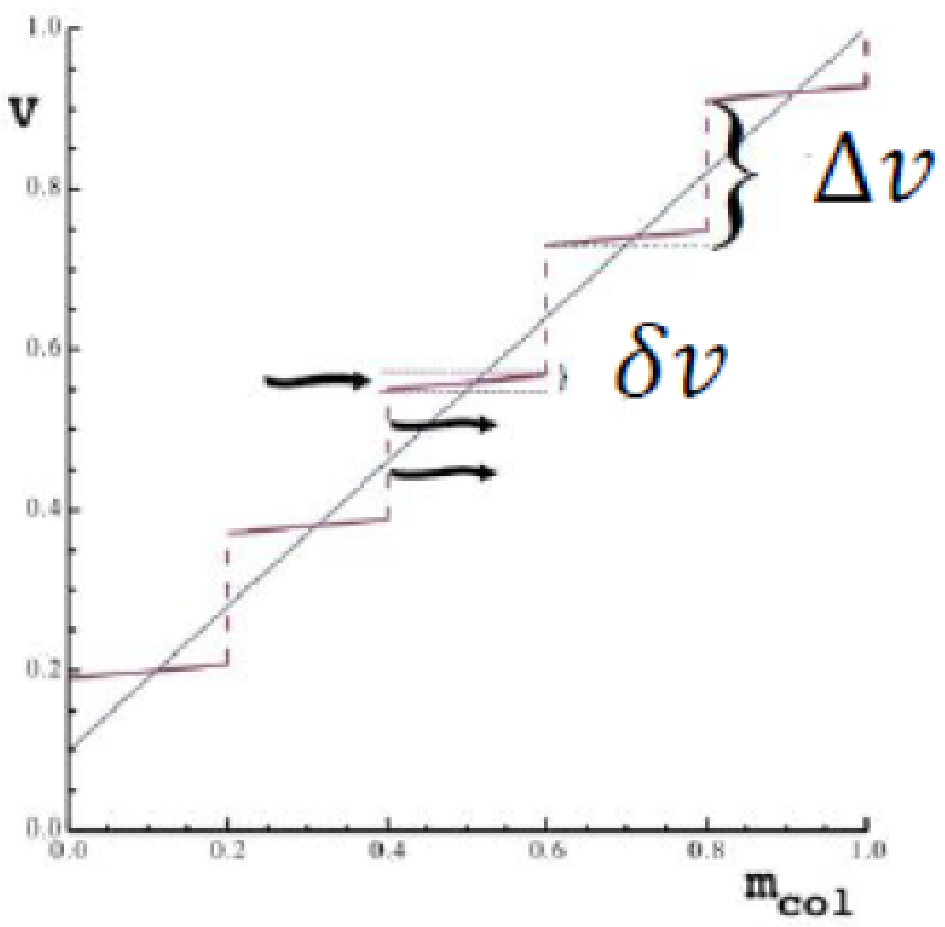}}
\end{minipage}
\caption{\textbf{Left, middle:} Randomly generated isotropic clumps of
  length scales $\ell_{\rm cl} \sim 0.1 r$ and velocity-stretch
  porosity with $h_\infty=0.5 R_\ast$ (left) and $h_\infty=5.0 R_\ast$
  (middle), that are lit from behind by a uniform source. Adapted from
  \citet{Sundqvist11b}. \textbf{Right:} Phenomenological velocity
  stair-case model vs. wind mass, where the dense clumps have velocity
  spans $\delta v$ that are separated by an amount $\Delta v$. Here
  the associated velocity filling factor is $\fvel = \delta v/ \Delta
  v = 1/10$ and the straight line represents a smooth CAK velocity
  law. Adapted from \citet{Owocki08}.}
\label{Fig:porvor}
\end{figure}

It is important to realize that the above results all rely on the
assertion that individual clumps are optically thin; if this is not
true for the investigated process, additional effects become important
in the radiation transfer. Namely, optically thick clumps lead to a
local self-shielding of opacity within the clumps, which in turn
allows for increased escape of radiation through porous channels in
between the clumps. The essential effect of such \textit{porosity} is
to reduce the medium's effective opacity as compared to an optically
thin clump model \citep{Feldmeier03, Owocki04}; Fig.~\ref{Fig:porvor}
graphically illustrates this general increase in transparency for
porous media. The trade-off between porosity and mass-loss rate thus
opposes the trade-off for density-squared diagnostics between
optically thin clumping and mass-loss rate , i.e. neglecting porosity
in cases where it is important can lead to \textit{underestimated}
mass-loss rates.

\begin{figure}[h]
\centering
\begin{minipage}{4.0cm}
\resizebox{\hsize}{!}
{\includegraphics[angle=90]{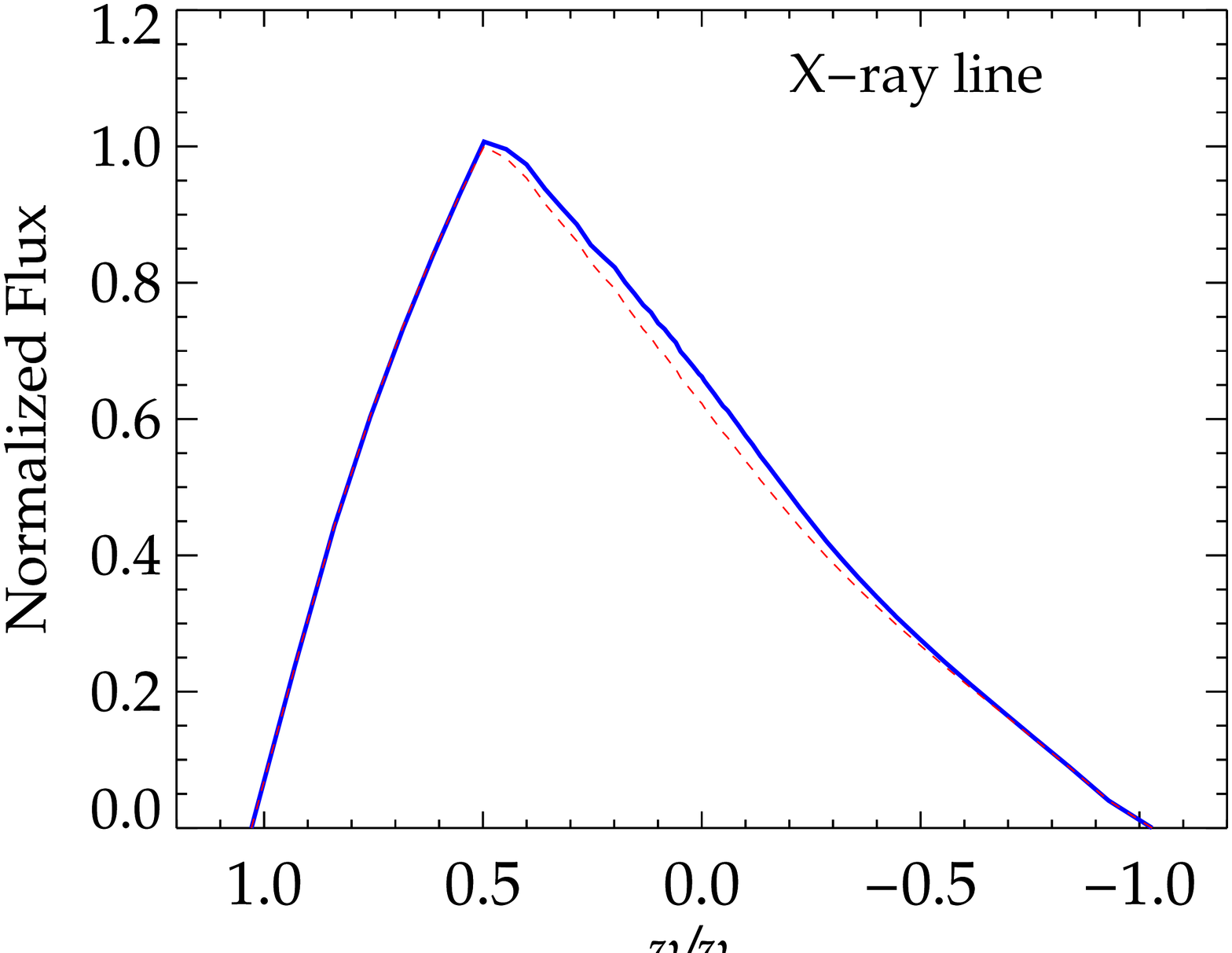}}
\end{minipage}
\begin{minipage}{4.0cm}
\resizebox{\hsize}{!}
{\includegraphics[angle=90]{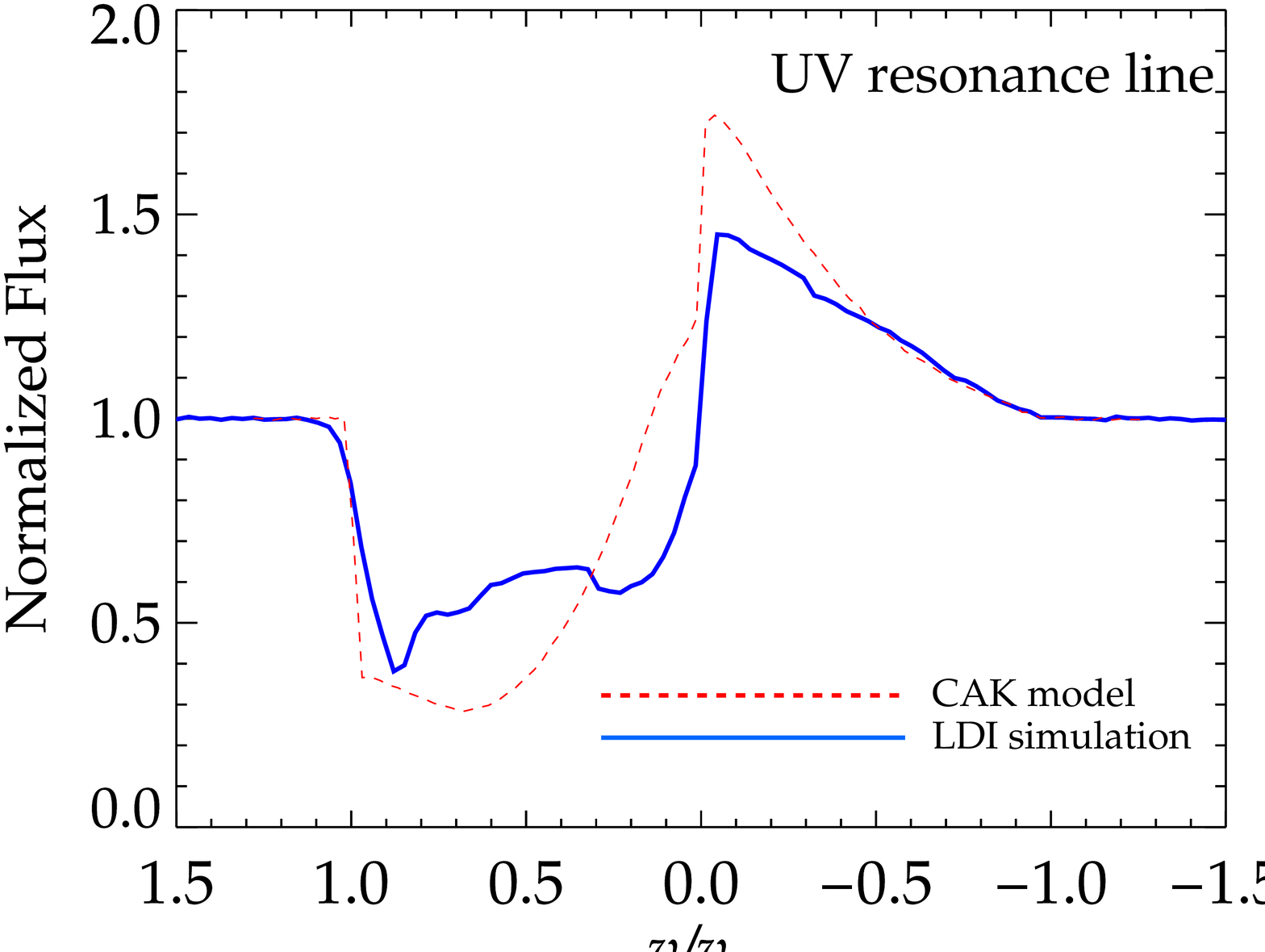}}
\end{minipage}
\begin{minipage}{4.0cm}
\resizebox{\hsize}{!}
{\includegraphics[angle=90]{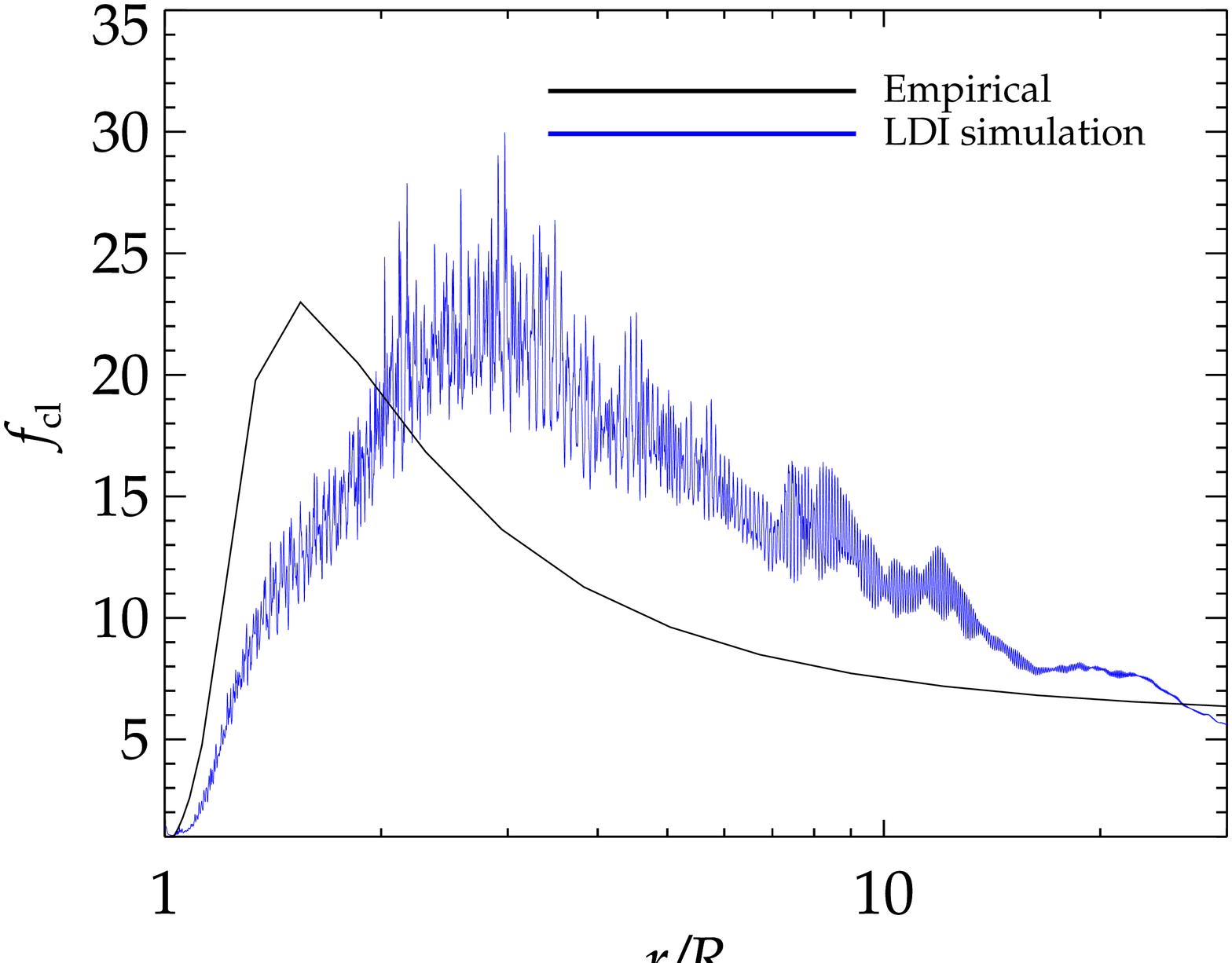}}
\end{minipage}
\caption{\textbf{Left, middle:} X-ray (left) and UV resonance (middle)
  line profiles for optical depths typical for $\zeta$ Pup at
  $\sim$\,15\,\AA~and for the \pv~doublet. Adapted from
  \citet{Sundqvist11b, Sundqvist11}. \textbf{Right:} Radial
  stratification of the clumping factor as derived for $\zeta$ Pup by
  \citet{Najarro11} (black) vs. theoretical predictions averaged over
  $\sim$\,1000 1-D LDI snapshots (blue). The same 1-D LDI simulations
  and synthesis calculation methods as in Fig.~\ref{Fig:LDI} are
  used.}
\label{Fig:fcl}
\end{figure}

\noindent \textbf{Porosity in continuum absorption of X-rays?}  For
\textit{continuum} processes, the key parameter controlling to which
extent porosity might be important is the so-called porosity length,
defined as the mean free path between clumps and also equal to the
clump length scale times the clumping factor, $h = \ell_{\rm cl}
\fcl$.  Simplified porosity wind models have been applied to
bound-free X-ray attenuation in O-star winds \citep{Feldmeier03,
  Oskinova04, Oskinova06, Owocki06}. The specific results of these
studies depend on whether isotropic or anisotropic clumps are assumed,
where the latter tends to enhance porosity effects \citep{Oskinova06,
  Sundqvist11b}. A key result for isotropic porosity is that it is
quite difficult to make individual clumps optically thick, simply
because the total integrated optical depths are quite low for X-ray
absorption in O stars \citep{Cohen10, Cohen11}. For `velocity stretch'
models in which the porosity length scales with the local wind
velocity, $h = h_\infty (v/v_\infty)$, terminal porosity lengths well
above a stellar radius typically are required for a significant effect
\citep{Cohen08, Sundqvist11b}. Therefore it seems rather unlikely that
porosity from LDI structure could have a significant effect on X-ray
line profiles; even though velocity stretching and clump-clump
collisions in 1-D LDI models can cause clump separations in the outer
wind to become on the order of a stellar radius, clump separations at
lower radii tend to be much smaller, on the order of the Sobolev
length $L_{\rm sob} = v_{\rm th}/(dv/dr) \approx (v_{\rm th}/v) R_\ast
\approx 0.01 R_\ast$. The left panel of Fig.~\ref{Fig:fcl}
demonstrates explicitly that X-ray line profiles with an opacity
computed directly from the LDI density structures described in
Sect.~\ref{LDI} are very similar to profiles computed using a smooth
CAK model, indicating small porosity effects. \\

\noindent \textbf{Vorosity in resonance line absorption?}  Contrasted
to the case of continuum absorption discussed thus far, clumps can
easily become optically thick in the inherently very strong UV
resonance lines, as first pointed out by \citet{Oskinova07}. In a
rapidly accelerating stellar wind, each \textit{line} photon can only
interact with the wind material within a very narrow spatial range,
due to the narrow Doppler width of the line profile. This makes it
possible for line photons to leak through the wind via `porous'
channels in velocity space, without ever interacting with the
optically thick clumps (Fig.~\ref{Fig:porvor}, rightmost panel); hence
this effect has been dubbed velocity porosity, or `vorosity'
\citep{Owocki08}. For a simple model with zero thermal speed, such
vorosity may be characterized using solely a \textit{velocity} filling
factor $\fvel \equiv \delta v/\Delta v$, as defined in
Fig.~\ref{Fig:porvor}. The basic point though, is that this photon
leakage in velocity space produces the same general effect as spatial
porosity, namely a reduction in effective opacity as compared to an
optically thin clump model.

\cite{Prinja10} found empirical evidence for such optically thick
clumping by analyzing the Si\,{\sc iv} resonance doublet in B
supergiants. As these resonance lines are unaffected by optically thin
clumps, the optical depth ratio between the blue and red line
components should reflect the underlying atomic physics, i.e.
$\tau_{\rm b}/\tau_{\rm r} = f_{\rm b}/f_{\rm r}=2.0$ where $f$ is the
oscillator strength, \textit{if} clumps were indeed optically
thin. However, the ratios inferred by \citet{Prinja10} spread over the
range $\sim$\,1 to 2, with a mean $\tau_{\rm b}/\tau_{\rm r} = 1.5 \pm
0.3$ derived from their core sample of 25 stars. This result is a
clear signature of optically thick clumps, and $\tau_{\rm b}/\tau_{\rm
  r} \approx 1.5$ has been shown to be consistent with calculations
using the analytic vorosity models by \citet{Sundqvist11}.

The fact that clumps seem to be optically thick in UV resonance lines
may then also help explain the unexpected weakness of observed UV
lines such as \pv~(Sect.~\ref{thin}), as demonstrated by first
attempts to include corresponding effects in line diagnostics
\citep{Oskinova07, Hillier08, Sundqvist10, Sundqvist11,
  Surlan11}. \citet{Sundqvist11} analyzed \pv~and \Ha~in the Galactic
O6 supergiant $\lambda$ Cep, accounting for structures in density as
well as for a non-monotonic velocity field. As \Ha~is rather
unaffected by optically thick clumping in this O star, whereas the
strengths of the \pv~lines are substantially reduced \citep[see
  also][for similar results for $\zeta$ Pup]{Oskinova07}, the derived
mass-loss rate is now only a factor of two lower than the prediction
by \citet{Vink00}. This significantly alleviates previous
discrepancies, and suggests that only mild reductions of current
theoretical mass-loss rates are necessary.

However, while the middle panel of Fig.~\ref{Fig:fcl} demonstrates
that vorosity is indeed present also in LDI simulations, the resulting
absorption reduction is generally less than needed to fully explain
the \pv~observations \citep[see also][]{Owocki08}. This may indicate
that the present generation of LDI simulations do not properly resolve
the internal clump velocity structures, and thus overpredict the clump
velocity spans $\delta v$. Moreover, another problem with these
simulations is that they do not show enough structure in the inner
wind to reproduce the observed \Ha~emission \citep{Sundqvist11}. We
conclude this review by briefly discussing this discrepancy between
predicted structure and observationally inferred clumping factors in
the inner wind.

\section{Clumping in the inner wind}
\label{stratification}

The right panel of Fig.~\ref{Fig:fcl} compares predictions from LDI
simulations with the radial stratification of \fcl~for $\zeta$ Pup, as
recently derived from the comprehensive multi-diagnostic study by
\citet{Najarro11}. While the agreement is good both in terms of
general behavior and in terms of absolute amount of clumping in the
outer wind, the observationally inferred clumping law peaks at much
lower radii than the theoretical one.

Considering the multitude of independent observational studies
indicating similar results \citep{Eversberg98, Bouret03, Puls06,
  Cohen11}, this motivates a re-investigation of the onset of wind
structure predicted by standard LDI simulations. Note that while the
LDI simulation displayed here does include base perturbations, it
still relies on the SSF formalism for calculating the line force and
the associated diffuse line drag (Sect.~\ref{LDI}). This probably
overestimates the damping of perturbations close to the wind base, as
suggested by first attempts to incorporate a more sophisticated escape
integral source function (EISF) into LDI models
\citep{Owocki99}. Along with suitable base perturbations, future work
will implement various forms of such EISFs into a new generation of
LDI simulations, and examine to what extent this may induce
substantial structure also near the wind base.

\acknowledgements J.O.S. gratefully acknowledge funding from NASA ATP
grant NNX11AC40G

\bibliographystyle{asp2010}
%\bibliography{sundqvist}
\bibliography{sundqvist}

\begin{thebibliography}{}
\expandafter\ifx\csname natexlab\endcsname\relax\def\natexlab#1{#1}\fi
\expandafter\ifx\csname url\endcsname\relax
  \def\url#1{\texttt{#1}}\fi
\expandafter\ifx\csname urlprefix\endcsname\relax\def\urlprefix{URL }\fi
\providecommand{\eprint}[2][]{\url{#2}}

\bibitem[{{Bouret} et~al.(2005){Bouret}, {Lanz}, \& {Hillier}}]{Bouret05}
{Bouret}, J.-C., {Lanz}, T., \& {Hillier}, D.~J. 2005, \aap, 438, 301

\bibitem[{{Bouret} et~al.(2003){Bouret}, {Lanz}, {Hillier}, {Heap}, {Hubeny},
  {Lennon}, {Smith}, \& {Evans}}]{Bouret03}
{Bouret}, J.-C., {Lanz}, T., {Hillier}, D.~J., {Heap}, S.~R., {Hubeny}, I.,
  {Lennon}, D.~J., {Smith}, L.~J., \& {Evans}, C.~J. 2003, \apj, 595, 1182

\bibitem[{{Carlberg}(1980)}]{Carlberg80}
{Carlberg}, R.~G. 1980, \apj, 241, 1131

\bibitem[{{Castor} et~al.(1975){Castor}, {Abbott}, \& {Klein}}]{Castor75}
{Castor}, J.~I., {Abbott}, D.~C., \& {Klein}, R.~I. 1975, \apj, 195

\bibitem[{{Cohen} et~al.(2011){Cohen}, {Gagn{\'e}}, {Leutenegger}, {MacArthur},
  {Wollman}, {Sundqvist}, {Fullerton}, \& {Owocki}}]{Cohen11}
{Cohen}, D.~H., {Gagn{\'e}}, M., {Leutenegger}, M.~A., {MacArthur}, J.~P.,
  {Wollman}, E.~E., {Sundqvist}, J.~O., {Fullerton}, A.~W., \& {Owocki}, S.~P.
  2011, \mnras, 415, 3354

\bibitem[{{Cohen} et~al.(2008){Cohen}, {Leutenegger}, \& {Townsend}}]{Cohen08}
{Cohen}, D.~H., {Leutenegger}, M.~A., \& {Townsend}, R.~H.~D. 2008, in Clumping
  in Hot-Star Winds, edited by {W.-R.~Hamann, A.~Feldmeier, \& L.~M.~Oskinova},
  209

\bibitem[{{Cohen} et~al.(2010){Cohen}, {Leutenegger}, {Wollman}, {Zsarg{\'o}},
  {Hillier}, {Townsend}, \& {Owocki}}]{Cohen10}
{Cohen}, D.~H., {Leutenegger}, M.~A., {Wollman}, E.~E., {Zsarg{\'o}}, J.,
  {Hillier}, D.~J., {Townsend}, R.~H.~D., \& {Owocki}, S.~P. 2010, \mnras, 405,
  2391

\bibitem[{{Crowther}(2007)}]{Crowther07}
{Crowther}, P.~A. 2007, \araa, 45, 177

\bibitem[{{Dessart} \& {Owocki}(2003)}]{Dessart03}
{Dessart}, L., \& {Owocki}, S.~P. 2003, \aap, 406, L1

\bibitem[{{Dessart} \& {Owocki}(2005{\natexlab{a}})}]{Dessart05}
--- 2005{\natexlab{a}}, \aap, 437, 657

\bibitem[{{Dessart} \& {Owocki}(2005{\natexlab{b}})}]{Dessart05b}
--- 2005{\natexlab{b}}, \aap, 432, 281

\bibitem[{{Eversberg} et~al.(1998){Eversberg}, {Lepine}, \&
  {Moffat}}]{Eversberg98}
{Eversberg}, T., {Lepine}, S., \& {Moffat}, A.~F.~J. 1998, \apj, 494, 799

\bibitem[{{Feldmeier}(1995)}]{Feldmeier95}
{Feldmeier}, A. 1995, \aap, 299, 523

\bibitem[{{Feldmeier} et~al.(2003){Feldmeier}, {Oskinova}, \&
  {Hamann}}]{Feldmeier03}
{Feldmeier}, A., {Oskinova}, L., \& {Hamann}, W.-R. 2003, \aap, 403, 217

\bibitem[{{Feldmeier} et~al.(1997){Feldmeier}, {Puls}, \&
  {Pauldrach}}]{Feldmeier97}
{Feldmeier}, A., {Puls}, J., \& {Pauldrach}, A.~W.~A. 1997, \aap, 322, 878

\bibitem[{{Fullerton} et~al.(2006){Fullerton}, {Massa}, \&
  {Prinja}}]{Fullerton06}
{Fullerton}, A.~W., {Massa}, D.~L., \& {Prinja}, R.~K. 2006, \apj, 637, 1025

\bibitem[{{Hamann}(1981)}]{Hamann81}
{Hamann}, W.-R. 1981, \aap, 93, 353

\bibitem[{{Hamann} et~al.(2008){Hamann}, {Feldmeier}, \& {Oskinova}}]{Hamann08}
{Hamann}, W.-R., {Feldmeier}, A., \& {Oskinova}, L.~M. (eds.) 2008, {Clumping
  in hot-star winds}

\bibitem[{{Hillier}(1991)}]{Hillier91}
{Hillier}, D.~J. 1991, \aap, 247, 455

\bibitem[{{Hillier}(2008)}]{Hillier08}
--- 2008, in Clumping in Hot-Star Winds, edited by {W.-R.~Hamann, A.~Feldmeier,
  \& L.~M.~Oskinova}, 93

\bibitem[{{L{\'e}pine} \& {Moffat}(1999)}]{Lepine99}
{L{\'e}pine}, S., \& {Moffat}, A.~F.~J. 1999, \apj, 514, 909

\bibitem[{{L{\'e}pine} \& {Moffat}(2008)}]{Lepine08}
--- 2008, \aj, 136, 548

\bibitem[{{Lucy}(1984)}]{Lucy84}
{Lucy}, L.~B. 1984, \apj, 284, 351

\bibitem[{{MacGregor} et~al.(1979){MacGregor}, {Hartmann}, \&
  {Raymond}}]{Macgregor79}
{MacGregor}, K.~B., {Hartmann}, L., \& {Raymond}, J.~C. 1979, \apj, 231, 514

\bibitem[{{Moffat} et~al.(1988){Moffat}, {Drissen}, {Lamontagne}, \&
  {Robert}}]{Moffat88}
{Moffat}, A.~F.~J., {Drissen}, L., {Lamontagne}, R., \& {Robert}, C. 1988,
  \apj, 334, 1038

\bibitem[{{Najarro} et~al.(2011){Najarro}, {Hanson}, \& {Puls}}]{Najarro11}
{Najarro}, F., {Hanson}, M.~M., \& {Puls}, J. 2011, ArXiv e-prints.
  \eprint{1108.5752}

\bibitem[{{Oskinova} et~al.(2004){Oskinova}, {Feldmeier}, \&
  {Hamann}}]{Oskinova04}
{Oskinova}, L.~M., {Feldmeier}, A., \& {Hamann}, W.-R. 2004, \aap, 422, 675

\bibitem[{{Oskinova} et~al.(2006){Oskinova}, {Feldmeier}, \&
  {Hamann}}]{Oskinova06}
--- 2006, \mnras, 372, 313

\bibitem[{{Oskinova} et~al.(2007){Oskinova}, {Hamann}, \&
  {Feldmeier}}]{Oskinova07}
{Oskinova}, L.~M., {Hamann}, W.-R., \& {Feldmeier}, A. 2007, \aap, 476, 1331

\bibitem[{{Owocki}(2008)}]{Owocki08}
{Owocki}, S.~P. 2008, in Clumping in Hot-Star Winds, edited by W.-R. {Hamann},
  A.~{Feldmeier}, \& L.~M. {Oskinova}, 121

\bibitem[{{Owocki} et~al.(1988){Owocki}, {Castor}, \& {Rybicki}}]{Owocki88}
{Owocki}, S.~P., {Castor}, J.~I., \& {Rybicki}, G.~B. 1988, \apj, 335, 914

\bibitem[{{Owocki} \& {Cohen}(2006)}]{Owocki06}
{Owocki}, S.~P., \& {Cohen}, D.~H. 2006, \apj, 648, 565

\bibitem[{{Owocki} et~al.(2004){Owocki}, {Gayley}, \& {Shaviv}}]{Owocki04}
{Owocki}, S.~P., {Gayley}, K.~G., \& {Shaviv}, N.~J. 2004, \apj, 616, 525

\bibitem[{{Owocki} \& {Puls}(1996)}]{Owocki96}
{Owocki}, S.~P., \& {Puls}, J. 1996, \apj, 462, 894

\bibitem[{{Owocki} \& {Puls}(1999)}]{Owocki99}
--- 1999, \apj, 510, 355

\bibitem[{{Owocki} \& {Rybicki}(1984)}]{Owocki84}
{Owocki}, S.~P., \& {Rybicki}, G.~B. 1984, \apj, 284, 337

\bibitem[{{Owocki} et~al.(2011){Owocki}, {Sundqvist}, {Cohen}, \&
  {Gayley}}]{Owocki11}
{Owocki}, S.~P., {Sundqvist}, J.~O., {Cohen}, D.~H.~D., \& {Gayley}, K.~G.
  2011, in Four decades of Research on Massive Stars, edited by {C.~Robert,
  N.~{St-Louis}, \& L.~{Drissen}}

\bibitem[{{Prinja} et~al.(2005){Prinja}, {Massa}, \& {Searle}}]{Prinja05}
{Prinja}, R.~K., {Massa}, D., \& {Searle}, S.~C. 2005, \aap, 430, L41

\bibitem[{{Prinja} \& {Massa}(2010)}]{Prinja10}
{Prinja}, R.~K., \& {Massa}, D.~L. 2010, \aap, 521, L55+

\bibitem[{{Puls} et~al.(2006){Puls}, {Markova}, {Scuderi}, {Stanghellini},
  {Taranova}, {Burnley}, \& {Howarth}}]{Puls06}
{Puls}, J., {Markova}, N., {Scuderi}, S., {Stanghellini}, C., {Taranova},
  O.~G., {Burnley}, A.~W., \& {Howarth}, I.~D. 2006, \aap, 454, 625

\bibitem[{{Puls} et~al.(2008){Puls}, {Vink}, \& {Najarro}}]{Puls08}
{Puls}, J., {Vink}, J.~S., \& {Najarro}, F. 2008, \aapr, 16, 209

\bibitem[{{Robert}(1994)}]{Robert94}
{Robert}, C. 1994, \apss, 221, 137

\bibitem[{{Runacres} \& {Owocki}(2002)}]{Runacres02}
{Runacres}, M.~C., \& {Owocki}, S.~P. 2002, \aap, 381, 1015

\bibitem[{{Runacres} \& {Owocki}(2005)}]{Runacres05}
--- 2005, \aap, 429, 323

\bibitem[{{Rybicki} et~al.(1990){Rybicki}, {Owocki}, \& {Castor}}]{Rybicki90}
{Rybicki}, G.~B., {Owocki}, S.~P., \& {Castor}, J.~I. 1990, \apj, 349, 274

\bibitem[{{Sundqvist} et~al.(2011{\natexlab{a}}){Sundqvist}, {Owocki}, {Cohen},
  {Leutenegger}, \& {Townsend}}]{Sundqvist11b}
{Sundqvist}, J.~O., {Owocki}, S., {Cohen}, D., {Leutenegger}, M.~A., \&
  {Townsend}, R. 2011{\natexlab{a}}, \mnras, submitted

\bibitem[{{Sundqvist} et~al.(2010){Sundqvist}, {Puls}, \&
  {Feldmeier}}]{Sundqvist10}
{Sundqvist}, J.~O., {Puls}, J., \& {Feldmeier}, A. 2010, \aap, 510, A11

\bibitem[{{Sundqvist} et~al.(2011{\natexlab{b}}){Sundqvist}, {Puls},
  {Feldmeier}, \& {Owocki}}]{Sundqvist11}
{Sundqvist}, J.~O., {Puls}, J., {Feldmeier}, A., \& {Owocki}, S.~P.
  2011{\natexlab{b}}, \aap, 528, A64

\bibitem[{{\u{S}urlan}(2011)}]{Surlan11}
{\u{S}urlan}, B. 2011, in Four decades of Research on Massive Stars, edited by
  {C.~Robert, N.~{St-Louis}, \& L.~{Drissen}}

\bibitem[{{Vink} et~al.(2000){Vink}, {de Koter}, \& {Lamers}}]{Vink00}
{Vink}, J.~S., {de Koter}, A., \& {Lamers}, H.~J.~G.~L.~M. 2000, \aap, 362, 295

\end{thebibliography}

\end{document}